\documentclass[11pt]{article}
\usepackage{cospar}
\usepackage{url}
\usepackage{txfonts} 

\usepackage{graphicx}
\usepackage[figuresright]{rotating}

\usepackage[sectionbib]{natbib} 
\title{BeppoSAX observations of rotation-powered pulsars}

\author{G. Cusumano\address{Istituto di Astrofisica Spaziale e
Fisica Cosmica (IASF) - Sezione di Palermo, CNR, Via Ugo La Malfa
153, I-90146, Palermo, Italy}}

\begin{document}
\maketitle

\begin{abstract}
   During its life, BeppoSAX observed several isolated
   pulsars, giving important
   contributions to the knowledge of the phenomenology and
   properties of the
   X-ray emission from this class of sources. In this paper I
   review the spectral and
   timing results obtained from BeppoSAX observations. In
   particular,
   results on three Crab-like (Crab, PSR B1509-58 and PSR B0540-69) and
   3 millisecond
   pulsars (PSR J0218+42, PSR B1821-24 and PSR B1937+21) are
   reported.
\end{abstract}





\section{INTRODUCTION}

BeppoSAX satellite operated between April 1996 and April 2001. Its
scientific payload included 4 Narrow Field Instruments (hereafter
NFIs; for a detailed description of the scientific payload see
Boella et al. 1997) covering nearly three orders of magnitude for
energy from 0.1 up to 300 keV. During its lifetime BeppoSAX
observed about 20 fields of isolated pulsars detecting the source
in more than half of them.


In this review, the most relevant spectral and timing results for the three
more intense Crab-like pulsars, Crab, PSR B1509-58 and PSR B0540-69, and three
millisecond pulsars, PSR J0218+42, PSR B1937+21 and PSR B1821-24 are reported.

\section{THE CRAB PULSAR}

The Crab pulsar (PSR B0531+21) is the only rotation powered pulsar
observed at almost every energy band of the electromagnetic
spectrum. Its pulse profile is characterized by a double peak
structure with a phase separation of 0.4, approximately aligned in
absolute phase over all wavelengths. The relative intensity,
height and width of the two peaks varies with energy: in
particular, the first peak (P1), dominant al low energies, becomes
smaller than the second one (P2) in the soft $\gamma$ rays. The
enhancement of the bridge between these peaks - hereafter named
inter-peak (Ip) region - as function of energy is also well
evident. The Crab pulsar was first detected in the X-ray band by
Frits et al. (1969) and by Bradt et al. (1969). The HEAO-2
satellite produced the first high resolution ($\sim$ 4") image of
the Crab nebula/pulsar in X-ray along with the phase profile
(Harden and Seward 1984). Pravdo et al. (1997) reported detailed
pulse-phase resolved spectral analysis of the pulsed emission
obtained with the PCA (5~-~60 keV) and HEXTE (16~-~250 keV)
instruments on board RossiXTE. They found systematic spectral
changes in the photon power law index as a function of pulse
phase. Recently, Kuiper et al. (2001) by using the high energy
$\gamma$-ray data from the CGRO satellite together with data
obtained at soft/hard X-ray energies from other observatory
obtained an exhaustive high-energy picture of the Crab pulsar from
0.1 keV up to 10 GeV.

 BeppoSAX observed the Crab Nebula and Pulsar several
times because this source was used for periodical calibration of
the NFIs. All these observations provide a high statistics data
set ($\sim$ 200 ks) that allowed to perform a very accurate timing
and spectral analysis over an energy range wider than two orders
of magnitude from
   about 0.1 up to 300 keV (see Massaro et al. 2000 for details).
Timing analysis for this source was performed producing phase
histograms for each NFI and each pointing. The values of P and
$\dot{P}$ were derived from the Jodrell Bank Crab Pulsar Monthly
Ephemeris (http://www.jb.man.ac.uk). The good signal to noise
ratio of the NFIs, together with the high statistics allowed  to
describe the pulse profiles over the entire band in very high
detail as shown in the top panels of Fig. 1. The morphology change
of the profile as function of the energy is striking, with the
second peak increasing from the soft range of the LECS (0.1-10
keV) up to the harder range of the PDS (15-300 keV).

 \begin{figure}
\centerline{\includegraphics[width=105mm,angle=90]{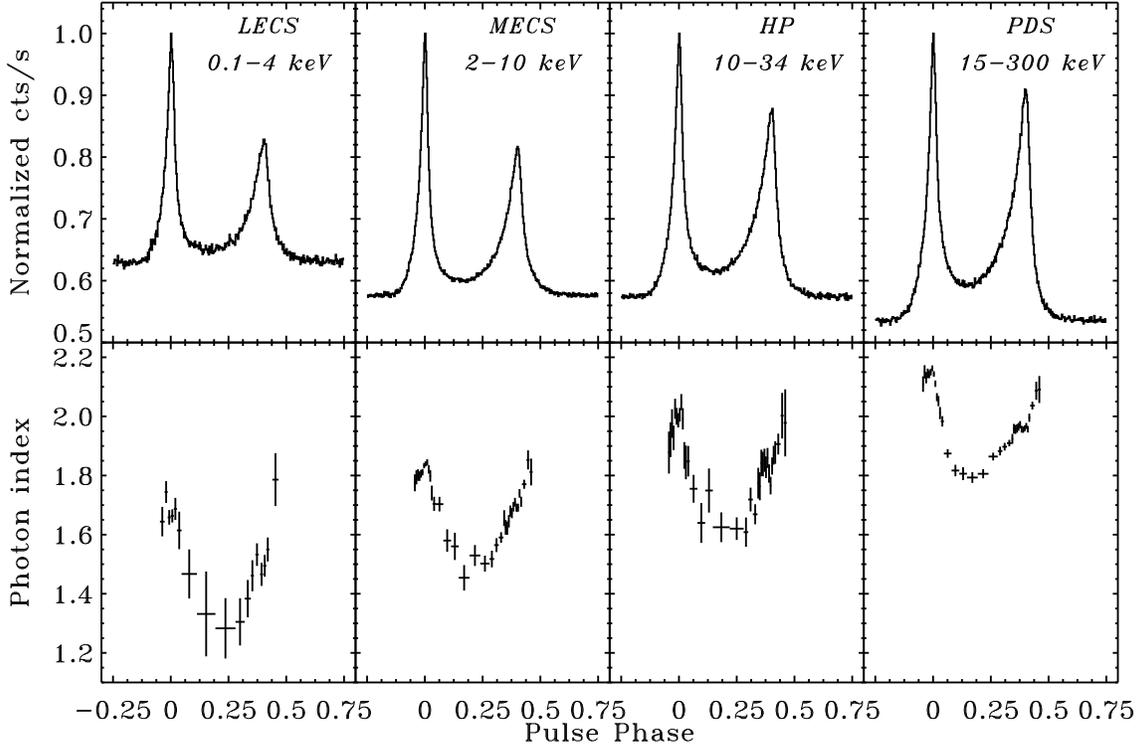}}
 \caption{Crab pulsar phase histograms ({\bf top}). Phase
 evolution of the spectral index ({\bf bottom}).}
 \end{figure}%

Phase resolved spectral analysis was performed with a minimum
phase resolution of 0.00667 unless the count number was not high,
as in the interpeak region and in some energy ranges, where wider
phase intervals were taken to reduce the statistical uncertainty
in estimating the spectral parameters. Modeling the spectra with a
simple power law, photon indices  were computed for each NFI. A
column density $N_H$=3.2$\times$10$^{21}$ cm$^{-2}$ derived from
the spectrum of the off-pulse interval (Massaro et al. 2000) was
fixed for all following fit procedures. The bottom panels in
Figure 1 show the phase evolution of the spectral index in the
four NFIs. The photon index changes along with the phase with a
similar phase dependence in each panel: The softer emission is in
P1 and P2 region while the hardest is in the middle of Ip; the
spectral index difference $\Delta \alpha$ is of the order of
0.3-0.5. Moreover, comparing the panels from left to right, we can
note that the photon index, relative to the same phase interval,
significantly increases with the energy. In particular, photon
index of P1 changes from 1.6 in the LECS range to 2.1 at higher
energies, that of P2 from 1.8 to 2.0 and Ip from 1.2 to 1.8. The
phase evolution of the spectral index is very similar to that
found by Pravdo et al (1997), but the spectral softening towards
higher energies is not so clear as in our results because of the
worse S/N ratio of the RXTE data at energies greater than 100 keV.

To explain the phase evolution of the spectral index and the
change of the pulse shape with the energy a two component model
was proposed by Massaro et al. (2000). They considered that the
observed pulsed emission is due to the superposition of two
components having different phase  and energy distributions. The
first component was assumed to have the same pulse profile as
observed at optical frequencies  with P1 much more prominent than
P2 and a very low intensity in the Ip region, the second one with
a harder spectrum and a phase profile  estimated from the BeppoSAX
data, by means of a fitting procedure of several pulse profiles at
different energies, in order to reproduce the observed pulse
profile when it was summed to the first component. This second
component was found to be much higher than the first one in the Ip
region, while the two components have a comparable intensity in
the P2 region.

As a consequence of the photon index behaviour with energy, shown
in Figure 1, we found that a single power law was not able to give
a satisfactory representation of the spectral distribution, for a
fixed phase interval region, over the entire BeppoSAX range
(0.1-300 keV). Better fits were obtained using a second order law
in the double log representation:

   $$ F(E)~=~K~(E/E_0)^{-(a+b~Log(E/E_0))}
\eqno(1)
$$

\noindent where $E_0$ is taken equal to 1 keV, and therefore $a$
corresponds to the photon index at this energy, while $b$ measures
the curvature of the spectral distribution. Applying this model to
three wider phase intervals, the first peak (P1) region
(0.99-0.01667), the Interpeak (Ip) region (0.01667-0.28) and the
second peak (P2) (0.3833-0.4167) we obtained a value for the
bending parameter $b$ of $\sim$ 0.15 over the three phase
intervals, and for the $a$ parameter the values 1.63$\pm$0.08
(P1), 1.31$\pm$0.03 (Ip) and 1.55$\pm$0.09 (P2). The best fit
parameters are summarized in Table 1.

\section{PSR~B1509-58}

PSR B1509$-$58 is one of the youngest rotation powered pulsars,
with a period of about 150 ms and a period derivative of
1.5$\times 10^{-12}$ s~s$^{-1}$, the highest spin-down rate of any
known pulsar. It was first  discovered in the soft X-rays by
Einstein (Seward \& Harnden 1982) and soon afterwards detected in
the radio band (Manchester, Tuohy \& D'Amico 1982). Later on, its
pulsed emission was also detected at hard X-ray and soft
$\gamma$-ray energies by several satellite and balloon-borne
experiments: EXOSAT (0.025--11 keV; Trussoni et al. 1990), Ginga
(2--60 keV; Kawai et al. 1992), SIGMA (40--300 keV; Laurent et al.
1994), Welcome-1 (94--240 keV; Gunji et al. 1994), ROSAT (0.1--2.4
keV; Greiveldinger et al. 1995), ASCA (0.5-10 keV, Saito 1998),
BATSE and OSSE aboard the Compton Gamma Ray Observatory (20-5000
keV; Wilson et al. 1993, Matz et al. 1994, Ulmer et al. 1993). A
detailed study of the pulse shape and spectrum in the energy range
2--200 keV with the PCA and HEXTE aboard the Rossi X-Ray Timing
Explorer has been presented by Marsden et al. (1997) and Rots et
al. (1998). Kuiper et al. (1999) reported definitive detection of
the pulsed signal in the COMPTEL band up to 10 MeV. Moreover, they
found a source compatible with the position of PSR~B1509$-$58 in
the skymaps of COMPTEL and EGRET between 10 and 100 MeV.

 \begin{figure}
\centerline{
\includegraphics[width=75mm,angle=-90]{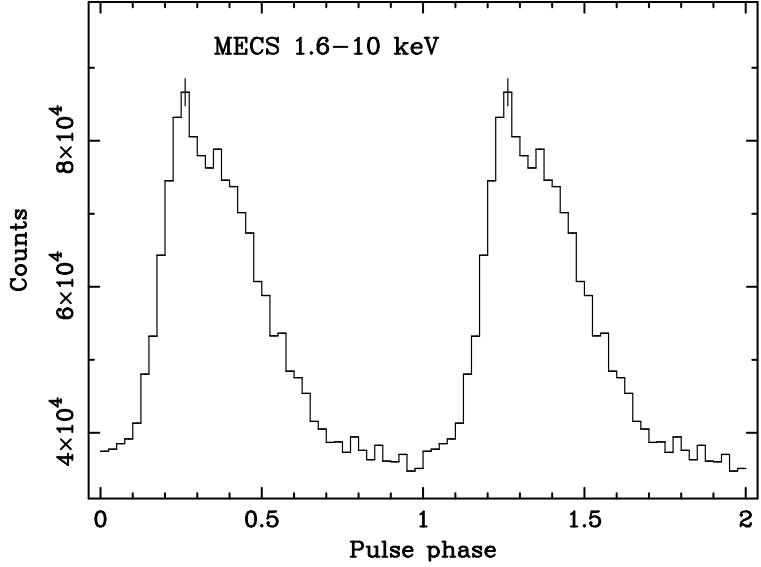}}
 \caption{Pulse profiles of PSR~B1509$-$58 in
the MECS energy band (1.6-10 keV). Two phase intervals are shown
for clarity. The 1 $\sigma$ uncertainty also is shown.}
 \end{figure}%

Given its high X-ray flux, second in intensity among the isolated
pulsars, the BeppoSAX observation allowed to perform timing and
spectral analysis at similar statistical significance as for the
Crab (Cusumano et al. 2001). The pulse profile extracted by
folding the MECS data (1.6-10 keV) is shown in Figure 2. The well
known single pulse shape is evident. It is markedly not symmetric
with the leading wing  much steeper than the trailing one. No
significant variation of the profile morphology has been detected
inside the BeppoSAX energy band (Cusumano et al. 2001). In the
spectral analysis, the pulsed and unpulsed spectra were extracted
for the phase intervals 0.17-0.53 and 0.77-1.07, respectively,
adopting the phase definitions of Mardsen et al. (1997). A column
density of (0.91$\pm$0.05) $\times 10^{22}$ cm$^{-2}$ was
evaluated by analyzing the LECS and MECS spectra of the off-pulse
region, assuming that the emission is mainly originating from the
nebula, at the same distance as the pulsar. The X-ray and soft
$\gamma$-ray pulsed spectra have usually been fitted with power
laws, but the resulting photon indices were generally found to
vary with the instrument energy windows. Trussoni et al. (1990)
first measured a photon index of about 1.1 with an EXOSAT
observation; the same value was found by Saito (1998) with ASCA.
Ginga and RXTE observations gave a photon index of 1.35 (Kawai et
al. 1993, Marsden et al. 1997), and an even steeper value of 1.68
was derived from BATSE and OSSE observations (Wilson et al. 1992,
Matz et al. 1994), suggesting a softening towards the higher
energies. However, no evidence for a spectral break was seen in
the RXTE data up to $\sim 200$ keV (Mardsen et al. 1997). Figure 3
reports the photon index obtained by fitting the pulsed spectra
for each of the BeppoSAX NFI with simple power laws. Notice that
the photon index for the PDS is significantly larger than that for
the MECS, while that derived for the LECS is smaller.

 \begin{figure}
\centerline{
\includegraphics[width=80mm,height=90mm]{cusumanofig3.ps}
\includegraphics[width=104mm,height=99mm]{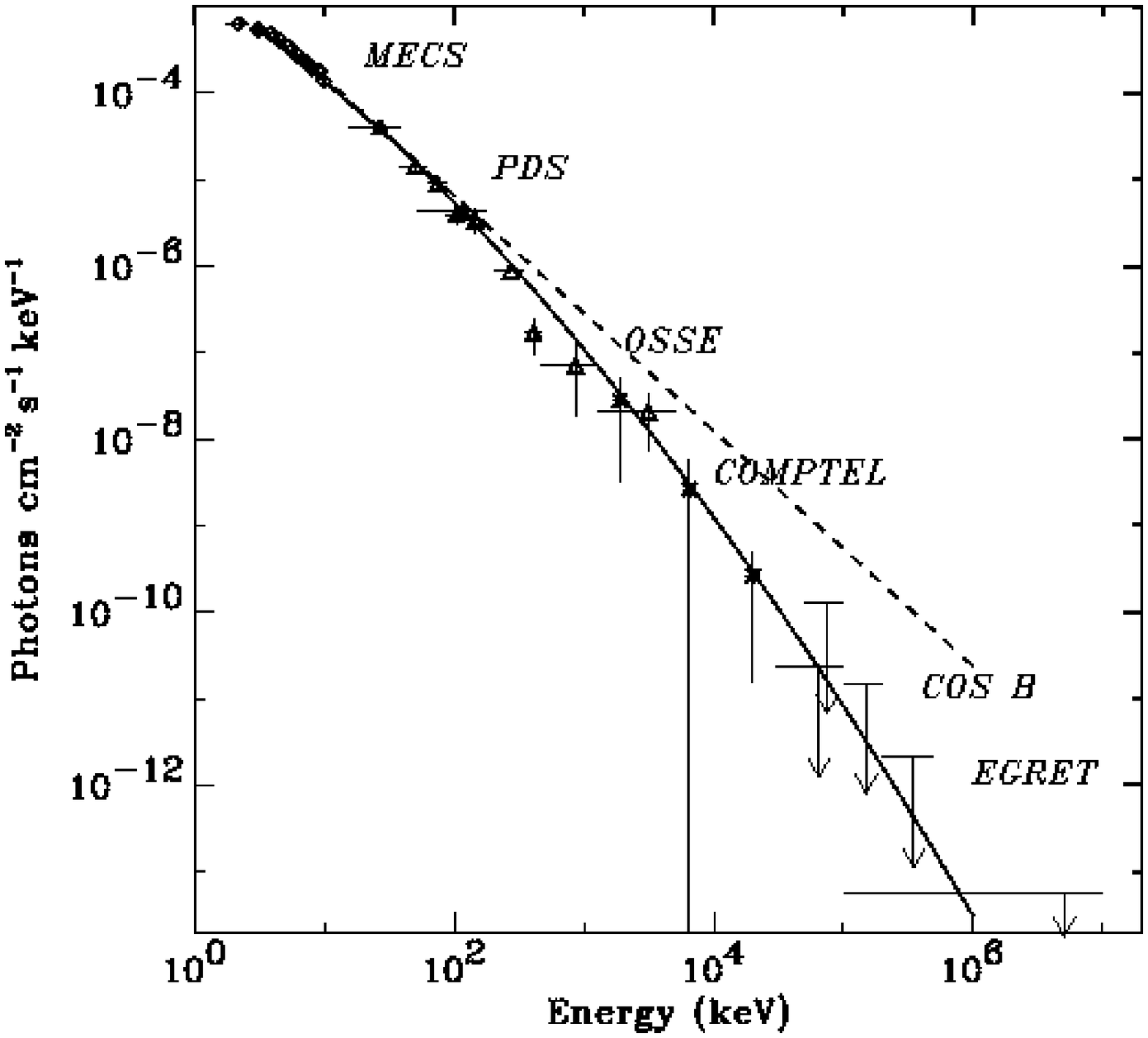}}
 \caption{{\bf Left}:
Photon indices of the pulsed emission of PSR B1509-58 obtained by
a power law fit to data of each of the NFIs. {\bf Right}:
PSR~B1509$-$58 deconvolved spectrum of the pulsed component.  Flux
values from BeppoSAX, OSSE, EGRET and COS B. The two lines
represent the best fit power--law (dashed) and best fit
energy--dependent power--law (solid) to the BeppoSAX data,
respectively.}
 \end{figure}%

In analogy with the Crab Pulsar, a simultaneous fit of a single power law shape to
the pulsed emission
over the entire BeppoSAX range did not give a satisfactory result
($\chi^2_{\nu}$ = 1.5 with 39 d.o.f.).
Fitting the whole BeppoSAX spectrum of PSR B1509-52
with the continuously steepening law of Eq. (1) gives a better representation
of the spectral distribution ($\chi^2_{\nu}$ = 0.74 with 38 d.o.f.).
The best fit values of the parameters
were: $a$~~0.96$\pm$0.08 and $b$~~0.16$\pm$0.04
(see Table 1 for a comparison with the Crab results).
The extrapolation of Eq. (1) to higher energies predicts fluxes in agreement
with the values measured by OSSE, BATSE and COMPTEL (Wilson et
al. 1993; Matz et al. 1993; Ulmer et al. 1993: Kuiper et
al. 1999).  This result is clearly shown in Figure 3.
For comparison, the extrapolation of the simple power--law model is also
shown.

\begin{table}
\vspace{-8mm}
\caption {Best fit spectral parameters of Crab pulsar, PSR B1509-58 and PSR B0540-69}
\begin{tabular}{lccccc}
\hline
\\
    & Crab pulsar & Crab pulsar & Crab pulsar & PSR B1509-58  & PSR B0540-69$^*$ \\
    &  (P1)       &      (Ip)   &   (P2)      &               &              \\
\hline
$a$ &1.63$\pm$0.08&1.31 $\pm$0.03&1.55$\pm$0.09&1.03$\pm$0.05&1.360$\pm$0.005 \\
$b$ &0.16$\pm$0.04&0.15 $\pm$0.09&0.13$\pm$0.03&0.13$\pm$0.02&0.143$\pm$0.003  \\
\hline
$^*$ de Plaa et al. (2003)
\end{tabular}
\end{table}

\section{PSR B0540-69}

 \begin{figure}
\centerline{ \includegraphics[width=95mm]{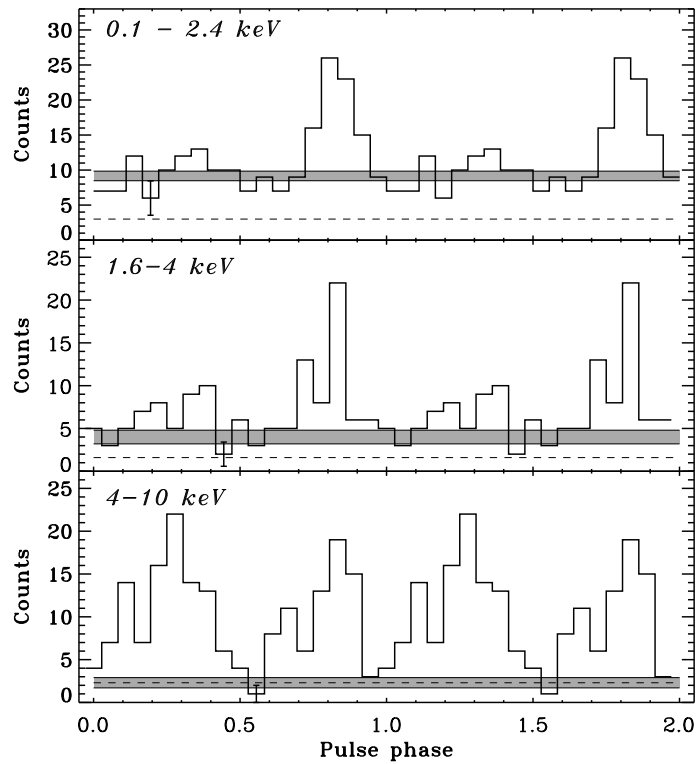}}
 \caption{PSR J0218+4232 phase histograms: energy range 0.1-2.4 keV
   from the
   ROSAT HRI (Kuiper et al. 1998); energy ranges 1.6-4 keV and
   4-10 keV from the
   BeppoSAX MECS.
   The shaded area represents the unpulsed level ($\pm$1
   $\sigma$) while the  ROSAT and MECS background levels from spatial analysis are
indicated with dashed lines.}
 \end{figure}%

Another Crab-like pulsar whose pulsed emission was detected by
BeppoSAX is PSR B0540-69. This pulsar,  located in the Large
Magellanic Cloud  has a pulsed period of about 50 ms and a large
period derivative of $4.79\times10^{-13}$ s s$^{-1}$, comparable
to that of the Crab pulsar.
 PSR B0540$-$69 was discovered in the
soft X-rays by Seward et al. (1984) with the Einstein Observatory.
Pulsations at optical frequencies were soon detected by
Middleditch and Pennypacker (1985) with a mean pulsed magnitude of
22.5. In the radio band PSR B0540$-$69  is a quite faint source
and pulsed signals were first observed only in 1989-90 (Manchester
et al. 1993). BeppoSAX detected the pulsar  only by the imaging
LECS and MECS instruments (Mineo et al. 1999 and Cusumano et al.
2003). Several estimates of the braking index of this pulsar have
been reported in the literature ranging from 2.01 $\pm$ 0.02
(Manchester and Peterson 1989, Nagase et al. 1990) to 2.74 $\pm$
0.10 (\"{O}gelman and Hasinger 1990). The BeppoSAX estimate of the
pulsar period, combined with earlier ASCA period evaluations,
allowed to measure a braking index of 2.10 $\pm$0.1. This results
has been recently confirmed by a combined timing analysis on an
extended data set including ASCA, BeppoSAX and RXTE observations
spanning a time interval of about 8 years (Cusumano et al. 2003).

Given the larger distance with respect to the previous two
pulsars, BeppoSAX data had not enough statistics to detected
pulsed emission at energies greater than 10 keV. The energy band
was not wide enough to investigate the presence of bending in the
spectral distribution like that found in the other two Crab-like
pulsars. Recently, de Plaa et al. (2003), making use of a long set
of RXTE observations derived the pulse profiles and spectral
distribution up to 50 keV. By fitting RXTE data together with the
ROSAT ones they found that pulsed emission can not be described by
a single power law because of the presence of a bend towards
higher energies. In analogy with the results shown just above, the
spectral distribution of this pulsar is well described by Eq. (1)
with a value of the bending parameter similar to that found for
Crab and PSR B1509-58 (Table 1).

\section{PSR J0218+4232}

PSR~J0218$+$42 is a 2.3 millisecond pulsar in a two day orbit
around a white dwarf companion (Navarro et al. 1995). The pulsar
has spin a down energy of $\dot{E}$=2.5$\times$10$^{35}$ erg
s$^{-1}$, a bipolar magnetic field component at the star surface
of B=4.3$\times$10$^{8}$ G and spin down age $\tau$ $\simeq$
4.6$\times$10$^{8}$ yr. Soft X-ray emission and pulsation were
first detected by ROSAT HRI (Verbunt et al. 1996; Kuiper et al.
1998) with a pulse profile characterized by a sharp main pulse and
an indication of a second peak. Emission at $\gamma$-rays from
this pulsar has been marginally detected with EGRET on board GRO
with a double peak profile with $\sim$ 3.5 $\sigma$ significance
(Kuiper et al. 2000). Recently, Chandra X-ray Observatory (CXO)
and Rossi X-ray Timing Explorer (RXTE) observations were able to
aligned the X-ray pulse profile with 2 of the 3 pulses visible at
radio-frequencies and with the two $\gamma$-ray pulses detected in
EGRET, increasing the significance of the $\gamma$ detection to a
4.9 $\sigma$ (Kuiper et al. 2002).

The BeppoSAX observation of the 2.3 millisecond pulsar PSR J0218+4232
provided for the first time detailed information on the
pulsar's temporal and spectral properties over the broad energy
band 1.6--10 keV (Mineo et al. 2000).
Pulsed emission was detected in the MECS energy band (1.6--10 keV) with a
significance of 6.8 $\sigma$ at the frequency extrapolated from radio
ephemeris.
Figure 4 shows the light curves in the energy bins 1.6-4 keV and 4-10 keV resulting
from folding all MECS events
with a phase resolution of 18 bins ($\sim$ 0.13 ms).
 The pulse profiles in the MECS ranges
are characterized by a double peak structure with
a relative phase separation of 0.47$\pm$0.05.
The MECS background level, determined from the spatial
analysis  is indicated with a dashed line.

In the same figure (top panel) the ROSAT (0.1--2.4 keV) profile
(Kuiper et al. 1998) is shown, shifted in phase to obtain the highest
peak coincident with the most
significant one in the MECS softer light curve (middle panel).

   \noindent

These profiles clearly show a change of the
relative peak intensities. The peak at phase 0.8 is
stronger in the low energy histograms
(top and middle panels), while that at phase 0.3, which is not
prominent in the ROSAT low energy phase histogram, becomes the
dominant feature above 5 keV (bottom panel). This behaviour reminds
the P1/P2 ratio observed in the Crab light curve at higher energies.
The unpulsed component determined by applying the bootstrap method
(Swanepoel et al. 1996) is shown as shaded area in all three panels of Figure 4.
Notice that, while in the ROSAT profile a DC component is
apparent above the background level, the same does not hold
for the 4--10 keV profile, where the background level is consistent
with the intensity measured in the valleys of the light curve.
In the intermediate energy profile there is evidence for
a DC component but its intensity is not as high as in the ROSAT profile.
Such effect can be interpreted as the presence of a quite soft
unpulsed emission.


The pulsed emission is well modeled with a power--law,
absorbed at low energy by the galactic column density with N$_H$
of 5$\times$10$^{20}$ cm$^{-2}$ (see Verbunt et al. 1996).
The measured photon index is 0.61$\pm$0.32 harder than for any other
isolated pulsar.
Spectra for the pulses P1 and P2 were also fitted  with
power-law models with spectral indices of $0.84\pm 0.35$
and $0.42\pm0.36$, respectively, in agreement with the trend seen in
Figure 5.
The unabsorbed (2--10 keV) pulsed  flux is 4.1$\times$10$^{-13}$
erg cm$^{-2}$ s$^{-1}$ implying a luminosity of $L_x =
1.3\times 10^{32} \,\Theta \; (d/5.7\;{\rm kpc})^2 $ erg s$^{-1}$,
where $\Theta$ is the solid angle spanned by the emission
beam.

\section{PSR~B1937+21}

PSR~B1937$+$21, the fastest known millisecond pulsar (P=1.558 ms),
is located  at a distance of 3.6 kpc as based on dispersion
measure (Taylor \& Cordes 1993). Its spin down energy is
$\dot{E}$=1.1$\times$10$^{36}$ erg s$^{-1}$ and the dipolar
magnetic field component at the star surface is
B=4.1$\times$10$^{8}$ G.
\\
It was discovered in radio by Backer et al. (1982).
 X--ray emission from this pulsar
 was first detected by ASCA (Takahashi et al. 2001) above 2 keV,
with a pulse profile characterized by a single sharply peak and a
pulsed fraction of 44\%. ASCA data allowed to aligned the X-ray
peak to the radio interpeak.

BeppoSAX detected pulsed emission from PSR~B1937+21 (Nicastro et al.
2002), the fastest known millisecond pulsar (1.55 ms).
Pulsed emission was detected in the MECS energy (1.6--10 keV)
with a significance of 11 $\sigma$ at a frequency deviating from
that extrapolated from the radio ephemeris by $\simeq -7.6 \times 10^{-6}$ Hz.
This discrepancy is consistent with a systematic
frequency error characteric for the timing precision
of the BeppoSAX clock for observation later than
2000 (Mineo et al. 2003).
%
Figure 5 shows the X-ray pulse profiles for the whole MECS range
(top panel) and for two energy sub-intervals 1.6--4 (middle panel)
and 4--10 keV (bottom panel). The pulse profile is characterized
by a double peak with a phase separation of $0.48\pm0.04$. The
significance of the second peak, first detected by BeppoSAX, is
about 5 $\sigma$. The comparison of the middle and bottom panels
in Figure 5, indicates a hint for variations in the ratio of the
two peaks. The unpulsed level determined by applying the bootstrap
method and the background level derived from a spatial analysis
are also shown in Figure 6. The DC level is compatible with the
background level implying that the  source is consistent with
being 100\% pulsed.

 \begin{figure}
\centerline{ \includegraphics[width=95mm]{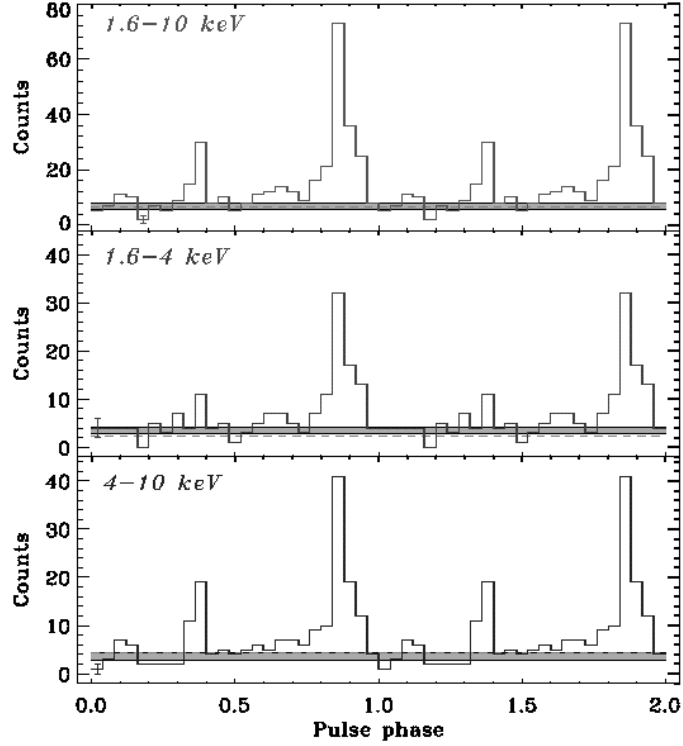}}
 \caption{PSR B1937+21 phase histograms:
1.6-10 keV (top panel), 1.6--4 (middle panel) and 4--10 keV
(bottom panel). The shaded areas indicate the DC level ($\pm 1
\sigma$), while the dashed lines indicate the background level.}
 \end{figure}%


The pulsed emission was modeled by an absorbed power law
with a photon index of 1.66$\pm$0.07 and $N_{\rm H}$ of
$2.3\times 10^{22}$ cm$^{-2}$.
The absorbed flux (2--10 keV) is $3.1\times 10^{-13}$
erg~cm$^{-2}$~s$^{-1}$
implying a luminosity of $L_X = 4.6\times 10^{31} \,
\Theta$ ($d/3.6$ kpc)$^2$
erg s$^{-1}$ and an X-ray efficiency of $\eta = 4\times
10^{-5}\, \Theta$.

\section {PSR B1821-24 }

PSR~B1821$-$24 is an isolated millisecond radio pulsar discovered
in the core of the globular cluster M28 (Lyne et al. 1987) with
high spin-down luminosity ($\dot{E}$=2.2$\times$10$^{36}$ erg
s$^{-1}$). It has a spin down period of P=3.05 ms and the dipolar
magnetic field component at the star surface is
B=2.2$\times$10$^{9}$ G. A marginal detection of X-ray pulsation
was first reported by Danner,  Kulkarni and Thorsett (1994) from a
ROSAT (PSPC) observation. A more significant detection was
performed by ASCA (Saito et al. 1997). The pulse profile is
characterized by a double peak with a phase separation of about
0.44. RXTE and Chandra observations allowed to aligned the
narrower X-ray peak to one of the radio peak (Rots et al. 1998;
Rutledge et al. 2003). Moreover, Chandra data provided detailed
spectral results on the pulsed emission (Becker et al. 2002).

BeppoSAX observation of PSR B1821-24 has provided more accurate
information on the  timing and spectral properties of the pulsar.
The pulsation in the MECS range (1.6 -- 10 keV) was detected at
high significance level (15 $\sigma$) at a frequency in agreement
with the radio ephemeris within the systematic deviation of the
onboard clock (Mineo et al. 2003). Figure 6 shows the X-ray pulse
profile in the total MECS range (top panel) and in two energy
sub-intervals 1.6--4 (middle panel) and 4--10 keV (bottom panel).
The shaded areas indicate the DC level ($\pm 1 \sigma$), while the
dashed lines indicate the background level. Note that the DC level
shown in the figure includes the contribution of the M28 globular
cluster in which the pulsar is located. Also for this millisecond
the pulse profile is characterized by a double peak structure with
phase separation of 0.45$\pm$0.03. A hint for variation in the
ratio of the two peaks with energy can be inferred comparing the
middle panel with the bottom panel of Figure 6.

The pulsed emission is well modeled with a power--law, absorbed at
low energy by the galactic column density with N$_H$ of
2.8$\times$10$^{20}$ cm$^{-2}$ and spectral index of
1.45$\pm$0.21. The unabsorbed (2--10 keV) pulsed  X-ray flux is
3.9$\times$10$^{-13}$ erg cm$^{-2}$ s$^{-1}$ implying a luminosity
of $L_x = 1.4\times 10^{33} \,\Theta \; (d/5.5\;{\rm kpc})^2 $ erg
s$^{-1}$.


 \begin{figure}
\centerline{ \includegraphics[width=95mm]{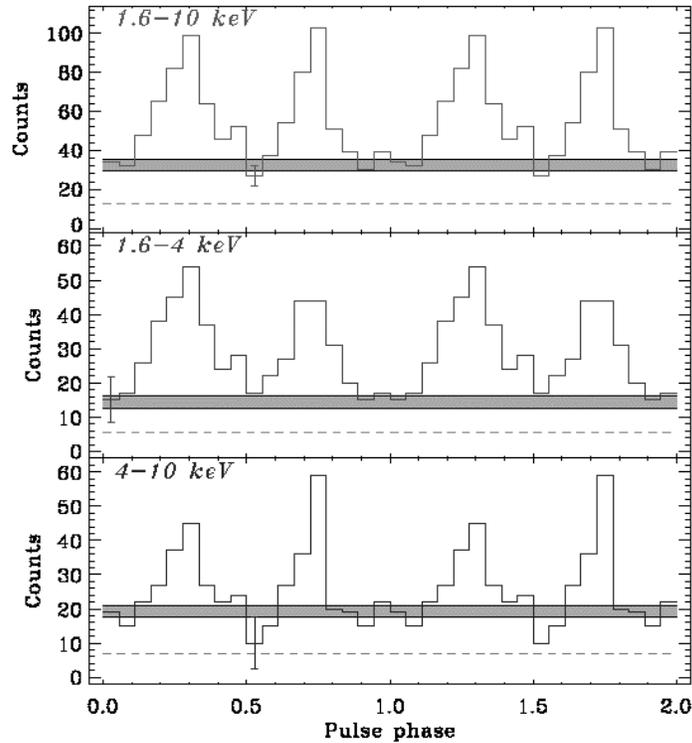}}
 \caption{PSR B1821-24 phase histogram:
1.6-10 keV (top panel), 1.6--4 (middle panel) and 4--10 keV
(bottom panel). The shaded areas indicate the DC level ($\pm 1
\sigma$), while the dashed lines indicate the background level.}
 \end{figure}%


\section {CONCLUSION}

On the basis of high quality BeppoSAX data, covering nearly three orders
of magnitude in energy (0.1-300 keV)
we  have showed that the energy spectra of the most intense Crab-like pulsars
can not be described by a single power lawn shape, but they show a significant steepening
towards higher energies. In fact, modeling the spectra of Crab,
finely selected in phase, with a simple power law we obtained photon indices
significantly increasing with the energy window of the NFIs. A similar result
was also obtained for PSR B1509-58.
Broad band spectra of three phase intervals of the Crab light
curve (P1,P2 and IP) and of the pulsed emission of PSR B1509-58
can be well modeled by a curved power law, Eq. (1), characterized
by a continuously bending of the spectral energy distribution.
Recently, De Plaa et al. (2003) found, analyzing  a set of RXTE
observations, that PSR B0540-69 shows a similar spectral
behaviour. The $b$ parameter, representing the curvature of the
spectral distribution, has been found to be similar, equal $\sim$
0.15 for all spectra. We stress that this new result has been
obtained just thanks to the broad band capability of BeppoSAX and
the excellent inter-calibration among the 4 Instruments.

Such curved spectra suggest the existence of a possible characteristic
energy, for instance the peak of the Spectral Energy Distribution (SED).
In the scenario of the outer gap model (Cheng et
al. 1986a, 1986b) the energy of the maximum of the SED
could be easily related, at least for this three Crab-like
pulsars, by a simple scaling law to some observable pulsar
parameters as the period and its first derivative (Cusumano
et al. 2001):

$$
(E_m/E_{mC}) \simeq (P/P_C)^{69/40} (\dot{P}/\dot{P}_C)^{11/40}
$$
$$
 \simeq (P/P_C)^{7/4} (\dot{P}/\dot{P}_C)^{1/4} \,\,.
\eqno(2)
$$

\noindent
where the index C refers to the Crab values.
The maximum of the SED has been calculated for the Crab to be 14 keV and
200 keV for the P1 and Ip region, respectively,
for PSRB1509-58 a higher value around 5 MeV
(Cusumano et al 2001), and for PSR B0540-69
the maximum is located around 170 keV.

The above scaling relation of the energy of the maximum of
the SED of PSR B1509-58
and PSR B0540-69 with respect the maximum for the Crab
matches quite well when the spectrum of the Crab
interpeak region is considered, while it is about one order
of magnitude lower for that of the first peak of Crab. This suggests a
similar mechanism of energy production for the X-ray emission of the
PSR B1509-58, PSR B0540-69 and for the Crab interpeak component that
correspond to the second component of the light curve in the model
proposed by Massaro et al. (2000).

Pulsed emission has been detected in the range 1.6-10 keV
for the 3 millisecond pulsars. All detected light curves
have double peak profiles with high pulsed fraction and
with a phase separation of about 0.5. PSR J0218+4232
shows spectral variation with phase: the relative
intensity of the two peaks varies with energy with a
behaviour that strongly recalls the Crab pulsar. Similar behaviour
seems also present in the other two milliseconds (PSR B1937+21  and
PSR B1821-24), although it needs to be confirmed with future
more significant data set.
The pulsed emission of these milliseconds is modeled by a flat power law
indicating a likely non-thermal emission.
The X-ray profile  morphology
and the hard non-thermal spectra suggest a similar magnetospheric
origin of the high-energy emission of these millisecond pulsars and the Crab.
The comparison with Crab is additionally enforced
by the similar values of the magnetic field at the light cylinder as
already pointed out by Kuiper et al. (1998) and Kawai and Saito (1999):
the magnetic field strength $B_L$ of PSR J0218+4232, PSR B1821-243
and PSR B1937+21 is in the range of 3 and 10 $\times$ 10$^5$ Gauss,
and that of the Crab is 9 $\times$ 10$^5$ Gauss.
On the other hand, the magnetic field strength at the neutron star surface
$B_S$  of the millisecond pulsars is more than 4 order of magnitude weaker
than the Crab one.

The similar value of $B_L$ seen in these millisecond pulsars and in the Crab
suggest that the magnetic field strength near the light cylinder is
a key parameter to explain their high energy emission,  indicating
as likely model for the high energy emission the outer gap model.
However, the polar cap scenario can also account for high-energy
emission from these milliseconds pulsars (Luo, Shibata and
Melrose 2000; Dyks and Rudak 2002).

\section*{ACKNOWLEDGEMENTS}

I am grateful to T. Mineo, E. Massaro and L. Nicastro for their
useful comments and their help for preparing this paper. The
author is very grateful to the referee, W. Hermsen, for the
comments and suggestions that improved the paper.

E-mail adress of G. Cusumano cusumano@pa.iasf.cnr.it
\\
\\

\end{document}